\documentclass[twocolumn,aps,superscriptaddress]{revtex4-2}
\usepackage{multirow}
\usepackage{float}
\usepackage{graphicx}
\usepackage{dcolumn}
\usepackage{bm}
\usepackage{bbold}
\usepackage{braket}
\usepackage{amssymb,amsmath}
\usepackage[section]{placeins}

\usepackage{color}
\usepackage{amsfonts}
\usepackage{subfigure}
\usepackage{array}
\usepackage{enumerate}

\usepackage{xspace}
\usepackage{siunitx}
\usepackage{xfrac}
\usepackage{hyperref}
\usepackage[nameinlink]{cleveref}
\usepackage{appendix}
\usepackage{subfigure}
\usepackage{xifthen}
\usepackage{xcolor}
\hypersetup{
	colorlinks,
	linkcolor={red!75!black},
	citecolor={blue!75!black},
	urlcolor={blue!75!black}
}
\graphicspath{{./figures/}{./}}
\begin{document}
\title{Nanohertz gravitational waves and primordial  quark nuggets from dense QCD matter in the early universe}
\author{Jingdong Shao}
\email[]{shaojingdong19@mails.ucas.ac.cn} 
\affiliation{School of Physical sciences, University of Chinese Academy of Sciences, Beijing 100049, China} 
\author{Hong Mao}
\email[]{mao@hznu.edu.cn}
\affiliation{School of Physics, Hangzhou Normal University, Hangzhou 311121, China}
\author{Mei Huang}
\email[]{huangmei@ucas.ac.cn~(corresponding author)}
\affiliation{School of Nuclear Science and Technology, University of Chinese Academy of Sciences, Beijing 100049, China}	
\begin{abstract}
The first-order QCD phase transition at high temperature features a large transition rate in the magnitude of $\beta/H \sim 10^4$ with induced stochastic gravitational waves typically lying in the LISA range. High baryon density in the early universe can be generated through Affleck-Dine baryogenesis. The baryon chemical potential enhances the potential barrier and significantly reduces the transition rate, which decreases from infinity at the critical end point (CEP) to zero at the critical nucleation point (CNP). Nanohertz gravitational waves can be produced in a narrow window of high baryon chemical potential with transition rate in the order of $\beta/H \sim 10^1$. When the phase transition rate reaches zero, the false vacuum of high baryon density quark matter is unlikely to decay and can persist over cosmological time scales. Therefore the primordial quark nuggets (PQN) can form and survive in the early universe as the seeds of compact stars, thereby dramatically accelerating the evolution of compact stars and the formation of galaxies, which may explain the high red-shift massive galaxies observed by the James Webb Space Telescope.
\end{abstract}

\maketitle
{\it Introduction:}
QCD phase transitions play important roles in the evolution of the early universe and compact stars. A first-order phase transition is required to produce stochastic background gravitational waves (SBGWs) in the early universe, while the peak structure of sound velocity describing neutron stars and GWs emitted from binary neutron stars may favor a crossover at high baryon density and zero chemical potential \cite{Huang:2022mqp}.

For physical quark mass, lattice QCD calculations indicate a crossover at high temperature \cite{Aoki:2006we} at zero or small chemical potential in 2-flavor as well as 3-flavor system \cite{Borsanyi:2020fev,Steinbrecher:2018phh}. The first-order QCD phase transition at high temperature can be found in a massless 3-flavor system \cite{Karsch:2001cy} as a chiral phase transition, the confinement-deconfinement phase transition in the pure gluon system \cite{Karsch:2001cy} and Friedberg-Lee model \cite{PhysRevD.15.1694,PhysRevD.16.1096,PhysRevD.18.2623}, as well as in the chirality imbalanced system \cite{Yu:2014sla,Shao:2022oqw}.

The QCD phase transition is also of first-order with high baryon density, which can be generated through the elegant and well established Affleck-Dine baryogenesis \cite{Affleck:1984fy,Linde:1985gh,Dine:2003ax} in the early universe and can be subsequently diluted through little inflation \cite{Borghini:2000yp,Boeckel:2009ej,Boeckel:2010bey,Schettler:2010dp,Boeckel:2011yj}. Or high baryon density clumps can be inhomogeneously distributed in the early universe, e.g., through first-order electroweak/QCD phase transitions or fluctuations, see Refs. \cite{Harrison:1968zza,Witten:1984rs,Applegate:1985qt,Orito:1996zb,PhysRev.167.1170,Megevand:2004ry,White:2021hwi}. The primordial quark nuggets (PQN) containing high baryonic number of the universe was proposed by Witten in 1984 \cite{Witten:1984rs} and has been investigated in many respects \cite{Applegate:1985qt,Schaeffer1986,Alam:1998xb,Horvath:2008qj}.

Recent observations from several pulsar timing array (PTA) collaborations, including the North American Nanohertz Observatory for Gravitational Waves (NANOGrav) \cite{NANOGrav:2023hde,Lerambert-Potin:2021ohy}, the Parkes Pulsar Timing Array (PPTA), the European Pulsar Timing Array (EPTA) and the Chinese Pulsar Timing Array (CPTA) \cite{Zic:2023gta,Reardon:2023gzh,EPTA:2023sfo,EPTA:2023fyk,Xu:2023wog}, have independently detected the evidence of stochastic GW signals in the nanohertz band. The nanohertz GWs may come from the orbiting or merger of supermassive (with mass $10^5-10^{10} M_{\odot}$) black hole binaries \cite{Sesana:2012ak}, or from cosmic phase transitions in the electroweak \cite{XIAO20233158} or QCD epoch \cite{Chen:2023bms}. 

Ref. \cite{Chen:2023bms} gives a bound for QCD phase transition rate $\beta/H<15$ to generate nanohertz GWs. The phase transition rate $\beta/H$, which describes the inverse duration of the phase transition, is a crucial parameter to decide the peak frequency and peak energy density of gravitational waves spectra. It is noticed that in some references, $\beta/H$ is treated as a free parameter. For example, $\beta/H$ is taken in the order of $1-10$ for pure gluon system to produce nanohertz GWs \cite{He:2022amv,He:2023ado}. However, for typical first-order chiral and confinement phase transitions at high temperature, the bona fide calculations in low-energy effective QCD-like theories and holographic QCD models give $\beta/H\sim 10^{4-5}$
\cite{Chen:2022cgj,Morgante:2022zvc,Helmboldt:2019pan,Han:2023znh,Gao:2024pqm}. Corresponding peak frequency of GWs typically lie in the region of $10^{-4}-10^{-2} {\rm Hz}$ with power spectrum in the range of $10^{-8}-10^{-7}$, which lies in the range of LISA and TAIJI. 

In this work, we investigate the possibility of producing nanohertz GWs from first-order QCD phase transitions, particularly if high baryon density QCD matter can be generated in the early universe. We will show that the baryon chemical potential can largely reduce the phase transition rate $\beta/H$, and there exists a narrow window of high baryon chemical potential with transition rate in the order of $\beta/H \sim 10^1$ to produce nanohertz GWs. There exists a critical quark chemical potential with zero phase transition rate $\beta/H=0$, which indicates the phase transition can barely start. Small $\beta/H$ means slow phase transitions and zero $\beta/H$ implies that the phase transition from quark matter to hadronic matter will never happen. It is possible that primordial "quarklet" or primordial quark nuggets exist in the early universe. 

{\it GWs from first-order phase transitions:}
First-order phase transitions complete via bubble nucleation. Once the phase transition starts, part of the universe tunnels to the true vacuum from the false vacuum, forming bubbles with lower vacuum energy density, then the latent heat released is converted into the energy of the bubble walls. These bubbles expand and collide and pass kinetic energy to surrounding media, generating GWs from the scalar field, the sound waves and the magnetohydrodynamic (MHD) turbulence \cite{Caprini:2015zlo}.

The bubble nucleation rate per Hubble volume per time has the exponential form $\Gamma(t)=Ae^{-S_4(t)}$ \cite{RN8,ellis2020gravitational,RN20}, 
where $S_4$ is the Euclidean action of an $O(4)$-symmetric solution and reduces to $S_3/T$ at high temperature $T$, and the coefficient $A$ has the form of $A(T)=T^4(S_3/2{\pi}T))^{3/2}$ \cite{RN15}.
Here $S_3$ is the bounce action of the tunneling path connecting the false vacuum and the true vacuum of a $O(3)$-symmetric bubble. For a scalar field $\Phi$ with potential $V(\Phi)$, the bounce action is
\begin{equation}
    S_3=\int d^3r(\frac{1}{2}(\nabla \Phi)^2+V)
\end{equation}
and the configuration of $\Phi(r)$ is decided by the $O(3)$-symmetric equation of motion
\begin{equation}
    \frac{d^2\Phi}{dr^2}+\frac{2}{r}\frac{d\Phi}{dr}=\frac{dV}{d\Phi}.
\end{equation}
with boundary conditions $\left.\frac{d\Phi}{dr}\right|_{r=0}=0$ and $\Phi(\infty)$ equal to the expectation value of $\Phi$ at the false vacuum. In specific models, the potential $V$ is usually the effective grand thermal dynamic potential $\Omega$.

Bubbles of true vacuum start to occur when the universe cools down to the nucleation temperature $T_n$, at which the nucleation rate catches the expansion rate of the Universe. In QCD epoch, $T_n$ can be quickly estimated by $S_3/T\sim 180$. More precisely, one bubble per Hubble volume per Hubble time ${\Gamma}(t)/H^4\sim1$ is expected at $T_n$ \cite{ellis2020gravitational,RN20}, where $H$ is the Hubble parameter given by the Friedmann equation.


Approximately $T_n$ is also the temperature of the thermal bath with weak reheating, thus the transition rate $\beta$ is defined as
\begin{equation}\frac{\beta}{H}=T_n\left.\frac{\mathrm{d}(\frac{S_3}{T})}{\mathrm{dT}}\right|_{T_n}.
\label{betaH}
\end{equation}

Another parameter which GWs spectra are sensitive to is $\alpha$, which measures the transition strength, i.e., relative magnitude between the latent heat released in the phase transition and the background radiation energy density $\rho_r$ \cite{RN15,ellis2020gravitational}. $\alpha$ can be calculated with finite $\mu$ as
\begin{equation}
\begin{aligned}
     \alpha & = \frac{-\Delta\rho+3\Delta p}{4\rho_r}
    \\
    & =\frac{1}{\rho_r}\left(\Delta p-\frac{T}{4}\left.\frac{\partial \Delta p}{\partial T}\right|_{T_p}-\frac{\mu}{4}\left.\frac{\partial \Delta p}{\partial \mu}\right|_{T_p}\right),
\end{aligned}
\end{equation}
where $\Delta$ means the difference between the true and false vacuum. $T_p$ is the percolation temperature and $T_p{\approx}T_n$ is used when $\beta/H \gg 1$, i.e., the false vacuum decays rapidly and hence the temperature is nearly constant during the phase transition. In specific models, the thermal background is not perfect ideal gas and thus the background radiation energy density $\rho_r=\frac{{\pi}^2gT^4}{30}$ ($g$ is the number of relativistic degrees of freedom) is not accurate, especially when the chemical potential $\mu$ can not be neglected compared with temperature $\mu\gtrapprox T$. Instead, the thermal energy density is given by the effective grand thermal potential $\Omega=-p$
\begin{equation}
    \rho=-(p-p_{vac})+\left.T_n\frac{\partial p}{\partial T}\right|_{T_n}+\mu \frac{\partial p}{\partial \mu},
\end{equation}
here $p_{vac}$ is the vacuum pressure at $T=\mu=0$ and must be deducted. 

The energy in the scalar field is negligibly small for relativistic bubbles \cite{RN69}, only two dominant sources, i.e., the sound wave and turbulence contribute to the total energy density GWs spectrum, i.e. 
\begin{equation}
h^2{\Omega}=h^2{\Omega}_{sw}+h^2{\Omega}_{tb}.
\label{gwsource}
\end{equation}In terms of the parameters above, the numerical results of GWs from sound waves and MHD turbulence take the forms of \cite{RN15,RN69}
\begin{equation}\label{11}h^2{\Omega}_{sw}(f)=2.65\times10^{-6}(\frac{H}{\beta})(\frac{\kappa_v\alpha}{1+\alpha})^2\left(\frac{100}{g}\right)^{\frac{1}{3}}v_wS_{sw}(f)\end{equation}
and
\begin{equation}\label{12}h^2{\Omega}_{tb}(f)=3.35\times10^{-4}(\frac{H}{\beta})(\frac{\kappa_{tb}\alpha}{1+\alpha})^2\left(\frac{100}{g}\right)^{\frac{1}{3}}v_wS_{tb}(f),\end{equation}
respectively. Parameters $\kappa_v$ and $\kappa_{tb}$ are respectively the fraction of the false vacuum energy converted into the kinetic energy of the plasma and the MHD turbulence which can be analytically fitted \cite{RN69,espinosa2010energy,kamionkowski1994gravitational}. GWs spectra are not sensitive to relativistic bubble velocity $v_w$ and thus in our following calculation we take a good approximati $v_w=\frac{\sqrt{1/3}+\sqrt{\alpha^2+2\alpha/3}}{1+\alpha}$ for strong phase transitions. $S_{sw}(f)$ and $S_{tb}(f)$ have the power-law form
\begin{equation}\label{13}S_{sw}(f)=\left(\frac{f}{f_{sw}}\right)^3\left(\frac{7}{4+3(\frac{f}{f_{sw}})^2}\right)^{\frac{7}{2}},\end{equation}
\begin{equation}\label{14}S_{tb}(f)=\left(\frac{f}{f_{tb}}\right)^3\left(1+\frac{f}{f_{tb}}\right)^{-\frac{11}{3}}\left(1+\frac{8{\pi}f}{h}\right)^{-1}.\end{equation}
The peak frequencies are
$f_{tb} = 1.42f_{sw}=16.36\frac{1}{v_w}\frac{\beta}{H}h$,  where $h=1.65\times10^{-6}\frac{T_n}{100{\rm GeV}}(\frac{g}{100})^{\frac{1}{6}} \mathrm{Hz}$ is the Hubble rate.

In the following, we investigate the GWs spectra induced by first-order deconfinement phase transition and chiral phase transition at high baryon chemical potentials by using two simple but representative models, i.e., the Friedberg-Lee (FL) Model and the quark-meson (QM) model, which can reveal the main features of GWs spectra induced by QCD phase transitions.

{\it Deconfinement phase transition in the Friedberg-Lee Model:} The Friedberg-Lee (FL) model provides a dynamical mechanism to confine quarks inside the nucleon by a complicated nonperturbative vacuum. It is described by the interaction of a phenomenological scalar field $\phi$ and quark field $\Psi$  \cite{PhysRevD.15.1694,PhysRevD.16.1096,PhysRevD.18.2623}, and the Lagrangian takes the form of
\begin{equation}
\mathcal{L}_{FL}=\bar\Psi(i\partial\!\!\!/-g \phi)\Psi+\frac{1}{2}\partial_\mu\phi\partial^\mu\phi-U_{FL}(\phi). \end{equation}
Here the potential $U_{FL}(\phi)$ takes a Ginzburg-Landau type with a quartic form
\begin{equation}
U_{FL}(\phi)=\frac{1}{2!}a\phi^2+\frac{1}{3!}b\phi^3+\frac{1}{4!}c\phi^4.  
\end{equation}
In the following numerical calculations, we fix  four parameters as $a=0.68921\mathrm{GeV}^2$, $b=-287.59\mathrm{GeV}$, $c=20000$ and $g=12.16$ as in Ref. \cite{Zhou_2021} to successfully reproduce the static properties of nucleon.   

Including one-loop contribution, the effective grand thermal potential at finite temperature and quark chemical potential is \cite{Quiros:1999jp,Laine:2016hma}
\begin{equation}
\begin{aligned}
& \Omega_{FL}=U_{FL}(\phi)+T\int \frac{d^3\vec p}{(2\pi)^3}\left\{{\rm ln} (1-e^{-E_{\phi}/T})\right.\\
&\left.-\nu\left[{\rm ln} (1+e^{-(E_{\Psi}-\mu)/T})+ {\rm ln} (1+e^{-(E_{\Psi}+\mu)/T})\right]\right\},
\end{aligned}
\end{equation}
with $2N_fN_c=2\times2\times3=12$ and $E_i=\sqrt{p^2+m_i^2}, (i=\phi,q)$. The effective mass of quark is $m_q=g\phi$ and that of the scalar field $m^2_{\phi}=a+b\phi+\frac{c}{2}\phi^2$, respectively. The order parameter $\phi$ can be determined by solving the gap equation
${\partial\Omega}_{FL}/{\partial\phi}=0$.

\begin{figure}[H]
    \centering
    \includegraphics[width=0.48\textwidth]{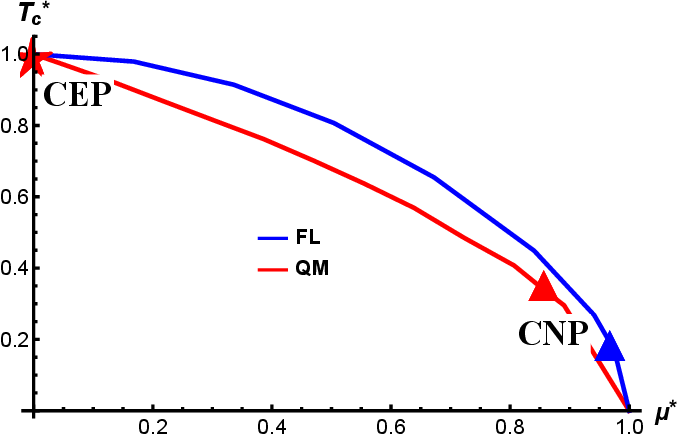}
    \caption{The first-order phase transition part in the scaled $T^*-\mu^*$ phase diagram of the FL model and QM model. The red star is for CEP at $\mu=299.4{\rm MeV}$ in the QM model, and triangle for CNP at $\mu=309.6{\rm MeV}$ in the QM model and $\mu= 287.55{\rm MeV}$ in the FL model, respectively. $T_c$ approaches $0$ at $\mu=311.3{\rm MeV}$ in the QM model and $\mu=297.5$ MeV in the FL model.}
    \label{tmuflandqm}
\end{figure}

The FL model can be used to describe a deconfinement phase transition. At $T<T_c$, there exists a soliton solution serving as a ``bag" to confine the quarks, while there is only a damping oscillation solution at $T>T_c$, thus the quarks are set free. It is noticed that the deconfinement phase transition in the FL model in the whole $T-\mu$ plane is of first-order. In Fig. \ref{tmuflandqm} we rescale the $T-\mu$ phase diagram to $T^*-\mu^*$ plane in the regime of $(0,1)$ with $T^*=T/T_{E}$ and $\mu^*=\mu/\mu_c$, where $T_E$ is the critical temperature at chemical potential $\mu=0$ in the FL model. 

{\it Chiral phase transition in the quark-meson model:}
The chiral phase transition can be described by the quark-meson (QM) model, and the Lagrangian of two-flavor QM model has the form of \cite{Gell-Mann:1960mvl,Wang:2023omt}
\begin{eqnarray}
 &&   \mathcal{L}_{QM}=\frac{1}{2}\partial^{\mu}\sigma\partial_{\mu}\sigma+\frac{1}{2}\partial^{\mu}\vec\pi\partial_{\mu}\vec\pi\nonumber \\
 &+& \bar\Psi i\partial\!\!\!/\Psi-g\bar\Psi(\sigma+i\gamma_5\vec\tau\cdot\vec\pi)\Psi-U_{QM}(\sigma,\vec\pi),
\end{eqnarray}
with the potential 
\begin{equation}
    U_{QM}(\sigma,\vec\pi)=\frac{\lambda}{4}(\sigma^2+\vec\pi^2-v^2)^2-H\sigma,
\end{equation}
where $\Psi=(u,d)$, and $\vec\tau$ the Pauli matrices. The interaction between quarks and scalar mesons including three pions $\vec\pi$ and one $\sigma$ meson.

The effective grand thermal potential in the QM model is
\begin{equation}
\begin{aligned}
&\Omega_{QM}=U_{QM}(\sigma,\vec\pi)-\nu\left\{\int\frac{d^3\vec p }{(2\pi)^3}E+\right.\\&\left.T\int\frac{d^3\vec p}{(2\pi)^3}\left[\mathrm{ln}(1+e^{-(E-\mu)/T})+\mathrm{ln}(1+e^{-(E+\mu)/T})\right]\right\}.
\end{aligned}
\end{equation}
with $E=\sqrt{\vec{p}^2+m_q^2}$.

The chiral symmetry is spontaneously broken in the vacuum and $\sigma$ obtains a non-zero vacuum expectation value  $\sigma=f_{\pi}=93\mathrm{\rm MeV}$, and three Goldstone Pions are massless in the chiral limit. The effective quark mass is $m_q=gf_{\pi}$ with $g=3.3$, where we have assumed that $m_q$ contributes $\frac{1}{3}$ mass of the nucleon. Partial conservation of the axial current gives the parameter $H=f_{\pi}m^2_{\pi}$ with pion mass $m_{\pi}=138\mathrm{MeV}$ in the case of nonzero current quark mass. The order parameter can be solved from the gap equation ${\partial\Omega}_{QM}/{\partial\sigma}=0$. 
The chiral transition in the QM model is a crossover in the low chemical potential region and of first-order at high baryon chemical potential region, thus there exists a CEP located at $T_E=32.17{\rm MeV}, \mu_E=299.4{\rm MeV}$. The first-order phase transition part starting from the CEP is shown in the rescaled $T^*-\mu^*$ phase diagram in Fig. \ref{tmuflandqm} with $T^*=T/T_{E}$ and $\mu^*=(\mu-\mu_{E})/(\mu_c-\mu_{E})$, where $T_E$ and $\mu_E$ are the critical temperature and chemical potential at CEP. When $T\rightarrow0$, the chiral symmetry restores at $\mu_c=311.3{\rm MeV}$.
 


\begin{figure}[H]
    \centering
    \includegraphics[width=0.48\textwidth]{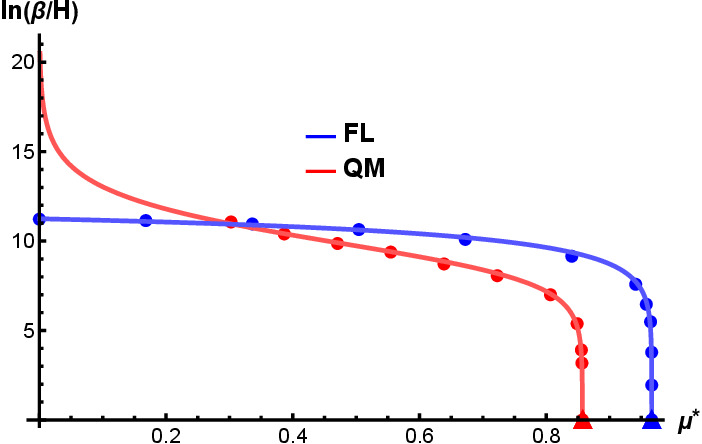}
    \caption{$\beta/H$ with different chemical potential $\mu$ in the FL (blue line) and QM (red line) model.}
    \label{betaflandqm}
\end{figure}

{\it The transition rate and GWs spectra:}
The transition rate $\beta/H$ for the first-order phase transition can be calculated through Eq. (\ref{betaH}), and the results of the FL model and QM model are shown in Fig. \ref{betaflandqm} with rescaled $\mu^*$. 

No CEP exists in the FL model. Values of $\beta/H$ decrease from $10^5$ to $10^3$ when $\mu$ increases from $0$ to $280$MeV and sharply falls from $10^3$ to $0$ in a very narrow window of the chemical potential. The point at which the nucleation barely starts is called the critical nucleation point (CNP), which corresponds to $\mu_{CNP}=287.55$ MeV in the FL model as marked by triangles in Fig. \ref{tmuflandqm} and \ref{betaflandqm}.

The CEP in the QM model is located at $T_E=32.17{\rm MeV}, \mu_E=299.4{\rm MeV}$ and it is found that the transition rate is almost infinity close to the CEP. Straightforward reason is that the potential barrier barely appears close to the CEP and thus the transition is ephemeral like a crossover. $\beta/H$ starts to fall from infinity from the CEP when chemical potential $\mu$ increases and soon reaches a plateau during which $\beta/H$ varies from $10^4$ to $10^3$ in the region of $309\mathrm{MeV}>\mu>306\mathrm{MeV}$ and sharply falls from $10^3$ to $0$ in a very narrow interval near CNP $\mu_{CNP}=309.6$MeV. The CNP at $\mu_{CNP}=309.6$MeV in the QM model is marked by triangles in Fig. \ref{tmuflandqm} and \ref{betaflandqm}.

\begin{figure}[H]
     \centering
     \includegraphics[width=0.95\linewidth]{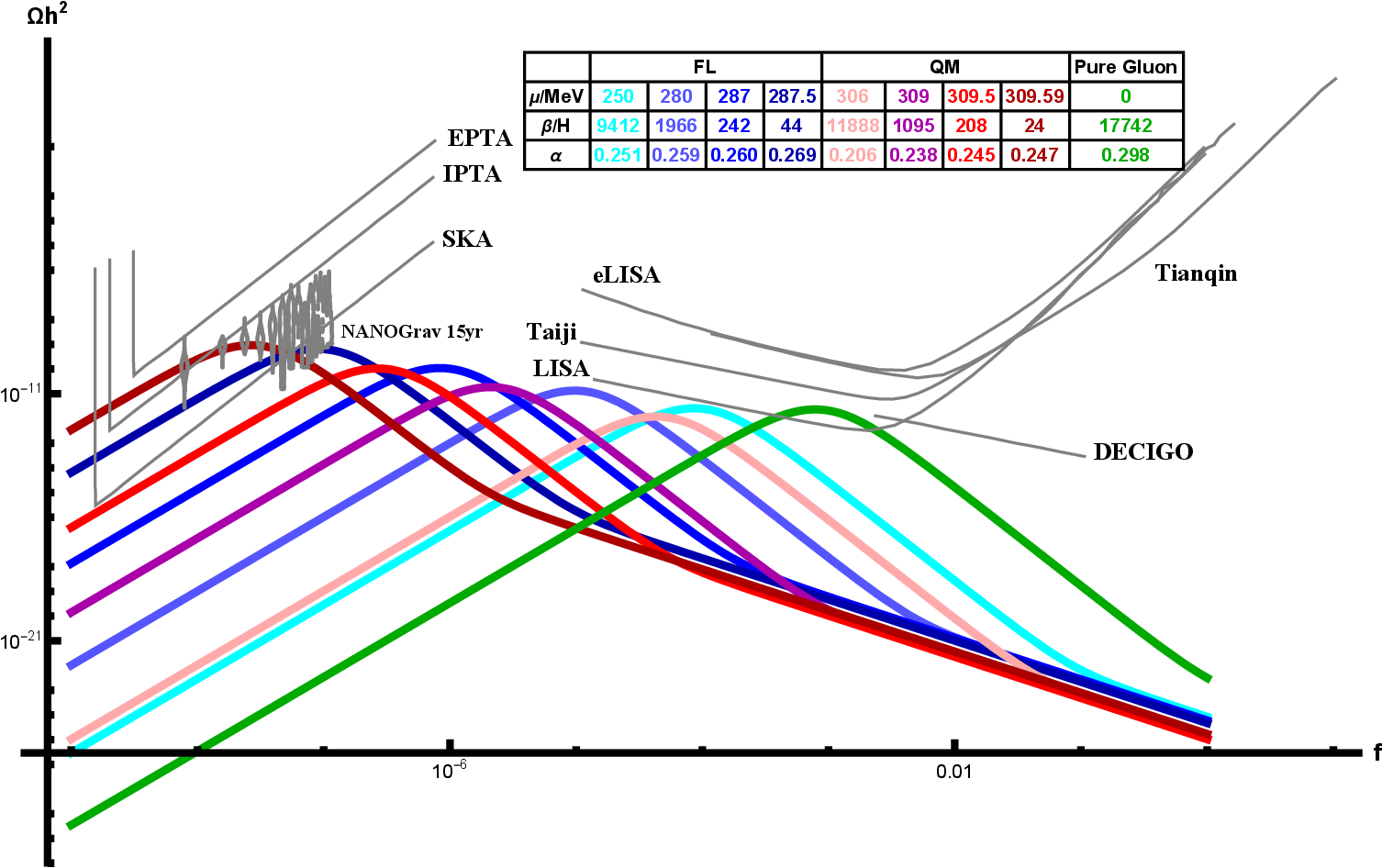}
     \caption{GWs spectra from first-order phase transitions in the FL model and QM model with different chemical potential. The pure gluon system at zero chemical potential is used as a reference.}
     \label{gw}
 \end{figure}

For GWs spectra, we select some specific values of chemical potential $\mu=306$, $309$, $309.5$, $309.59$ MeV in the QM model and $\mu=250$, $280$, $287$, $287.5$ MeV in the FL model, which corresponds to the magnitude of $\beta/H$ in the order of $10^4$, $10^3$, $10^2$, $10$, and the phase transition strength $\alpha$ is also calculated. 
These parameters are shown in the table inserted in Fig. \ref{gw} as well as GWs spectra, and the pure gluon system at zero chemical potential \cite{Shaojd2024long} is used as a reference.

It is noticed that in the pure gluon system without quark chemical potential, we have $\beta/H\sim 10^4$ and $\alpha\sim 0.3$, and the produced GWs fall in the LISA detection window. With chemical potential increases, the transition rate in the FL model and QM model decreases significantly, the produced GWs move to lower frequency. At high baryon density region with $\mu_B/T\sim 10$, the transition rate $\beta/H$ can reach the order of $10^1$, the produced nanohertz GWs can be detected by SKA, IPTA and EPTA, coinciding with NanoGrav data.

{\it Summary and Discussion:} In this work, we investigate the possibility of producing nanohertz GWs spectra from first-order deconfinement and chiral phase transition in the FL model and QM model at high baryon density. It has been shown by many effective QCD model calculations that the first-order QCD phase transitions at high temperatures is characterized by a large transition rate with the magnitude of $\beta/H \sim 10^4$, and its induced GWs typically fall in the LISA detection window. 

Through our studies, it is found that the baryon chemical potential enhances the potential barrier between the false and true vacuum, and can significantly reduces the transition rate. The transition rate is infinity at the CEP and drops to zero at CNP, which is one of the new finding features revealed in this study for the first time. The nanohertz GWs can be produced in a narrow window of high baryon chemical potential when the transition rate is in the order of $\beta/H \sim 10^1$. When the transition rate reaches zero, the false vacuum of quark matter is unlikely to decay and can persist over cosmological time scales, which supports the formation of primordial quark nuggets (PQN) proposed by Witten in 1984. Furthermore, The PQN will become the seeds of compact stars, dramatically accelerating the evolution of compact stars as well as the formation of galaxies. This may explain the high red-shift massive galaxies observed by the James Webb Space Telescope. Our results can be extended straightforwardly to high density of QCD-like dark matter \cite{Bai:2018dxf}. While QCD phase transitions can not be expected to complete at extreme super cooling as electroweak phase transitions to produce nanohertz GWs \cite{PhysRevLett.132.221001,SALVIO2024138639,XIAO20233158,2017EPJC77570K,Jinno2019GravitationalWF} due to temperature constrains from modern cosmology such as BBN, these chronical dense nuggets can survive to low temperature at which vacuum decay prevails thermal decay, the corresponding phase transition will eventually happen due to quantum effect at an extremely slow rate and probably eject part of matter or emit radiation.

It is worthy of mentioning that considering the diquark condensation, i.e., the color superconducting phase at high baryon density \cite{Rajagopal:2000wf,Huang:2004ik,Alford:2007xm}, the first-order phase transition line at high $\mu$ and low temperature might be changed. Furthermore, the effect from the primordial magnetic field in the QCD phase transition should also be taken into account in the future \cite{Li:2023yaj}.

{\it Acknowledgments:}
We thank F. Gao and J. Schaffner-Bielich for helpful discussions. This work is supported in part by the National Natural Science Foundation of China (NSFC) Grant Nos. 12235016 and 12221005, the Strategic Priority Research Program of Chinese Academy of Sciences under Grant No. XDB34030000.

\bibliographystyle{unsrt}
\bibliography{reference.bib}

\begin{thebibliography}{10}

\bibitem{Huang:2022mqp}
Yong-Jia Huang, Luca Baiotti, Toru Kojo, Kentaro Takami, Hajime Sotani, Hajime
  Togashi, Tetsuo Hatsuda, Shigehiro Nagataki, and Yi-Zhong Fan.
\newblock {Merger and Postmerger of Binary Neutron Stars with a Quark-Hadron
  Crossover Equation of State}.
\newblock {\em Phys. Rev. Lett.}, 129(18):181101, 2022.

\bibitem{Aoki:2006we}
Y.~Aoki, G.~Endrodi, Z.~Fodor, S.~D. Katz, and K.~K. Szabo.
\newblock {The Order of the quantum chromodynamics transition predicted by the
  standard model of particle physics}.
\newblock {\em Nature}, 443:675--678, 2006.

\bibitem{Borsanyi:2020fev}
Szabolcs Borsanyi, Zoltan Fodor, Jana~N. Guenther, Ruben Kara, Sandor~D. Katz,
  Paolo Parotto, Attila Pasztor, Claudia Ratti, and Kalman~K. Szabo.
\newblock {QCD Crossover at Finite Chemical Potential from Lattice
  Simulations}.
\newblock {\em Phys. Rev. Lett.}, 125(5):052001, 2020.

\bibitem{Steinbrecher:2018phh}
Patrick Steinbrecher.
\newblock {The QCD crossover at zero and non-zero baryon densities from Lattice
  QCD}.
\newblock {\em Nucl. Phys. A}, 982:847--850, 2019.

\bibitem{Karsch:2001cy}
F.~Karsch.
\newblock {Lattice QCD at high temperature and density}.
\newblock {\em Lect. Notes Phys.}, 583:209--249, 2002.

\bibitem{PhysRevD.15.1694}
R.~Friedberg and T.~D. Lee.
\newblock Fermion-field nontopological solitons.
\newblock {\em Phys. Rev. D}, 15:1694--1711, Mar 1977.

\bibitem{PhysRevD.16.1096}
R.~Friedberg and T.~D. Lee.
\newblock Fermion-field nontopological solitons. ii. models for hadrons.
\newblock {\em Phys. Rev. D}, 16:1096--1118, Aug 1977.

\bibitem{PhysRevD.18.2623}
R.~Friedberg and T.~D. Lee.
\newblock Quantum chromodynamics and the soliton model of hadrons.
\newblock {\em Phys. Rev. D}, 18:2623--2631, Oct 1978.

\bibitem{Yu:2014sla}
Lang Yu, Hao Liu, and Mei Huang.
\newblock {Spontaneous generation of local CP violation and inverse magnetic
  catalysis}.
\newblock {\em Phys. Rev. D}, 90(7):074009, 2014.

\bibitem{Shao:2022oqw}
Jingdong Shao and Mei Huang.
\newblock {Gravitational waves and primordial black holes from chirality
  imbalanced QCD first-order phase transition with P and CP violation}.
\newblock {\em Phys. Rev. D}, 107(4):043011, 2023.

\bibitem{Affleck:1984fy}
Ian Affleck and Michael Dine.
\newblock {A New Mechanism for Baryogenesis}.
\newblock {\em Nucl. Phys. B}, 249:361--380, 1985.

\bibitem{Linde:1985gh}
Andrei~D. Linde.
\newblock {The New Mechanism of Baryogenesis and the Inflationary Universe}.
\newblock {\em Phys. Lett. B}, 160:243--248, 1985.

\bibitem{Dine:2003ax}
Michael Dine and Alexander Kusenko.
\newblock {The Origin of the matter - antimatter asymmetry}.
\newblock {\em Rev. Mod. Phys.}, 76:1, 2003.

\bibitem{Borghini:2000yp}
N.~Borghini, W.~N. Cottingham, and R.~Vinh~Mau.
\newblock {Possible cosmological implications of the quark hadron phase
  transition}.
\newblock {\em J. Phys. G}, 26:771, 2000.

\bibitem{Boeckel:2009ej}
Tillmann Boeckel and Jurgen Schaffner-Bielich.
\newblock {A little inflation in the early universe at the QCD phase
  transition}.
\newblock {\em Phys. Rev. Lett.}, 105:041301, 2010.
\newblock [Erratum: Phys.Rev.Lett. 106, 069901 (2011)].

\bibitem{Boeckel:2010bey}
Tillmann Boeckel, Simon Schettler, and Jurgen Schaffner-Bielich.
\newblock {The Cosmological QCD Phase Transition Revisited}.
\newblock {\em Prog. Part. Nucl. Phys.}, 66:266--270, 2011.

\bibitem{Schettler:2010dp}
Simon Schettler, Tillmann Boeckel, and Jurgen Schaffner-Bielich.
\newblock {Imprints of the QCD Phase Transition on the Spectrum of
  Gravitational Waves}.
\newblock {\em Phys. Rev. D}, 83:064030, 2011.

\bibitem{Boeckel:2011yj}
Tillmann Boeckel and Jurgen Schaffner-Bielich.
\newblock {A little inflation at the cosmological QCD phase transition}.
\newblock {\em Phys. Rev. D}, 85:103506, 2012.

\bibitem{Harrison:1968zza}
E.~R. Harrison.
\newblock {Baryon Inhomogeneity in the Early Universe}.
\newblock {\em Phys. Rev.}, 167:1170--1175, 1968.

\bibitem{Witten:1984rs}
Edward Witten.
\newblock {Cosmic Separation of Phases}.
\newblock {\em Phys. Rev. D}, 30:272--285, 1984.

\bibitem{Applegate:1985qt}
J.~H. Applegate and C.~J. Hogan.
\newblock {Relics of Cosmic Quark Condensation}.
\newblock {\em Phys. Rev. D}, 31:3037--3045, 1985.

\bibitem{Orito:1996zb}
M.~Orito, T.~Kajino, R.~N. Boyd, and G.~J. Mathews.
\newblock {Geometrical effects of baryon density inhomogeneities on primordial
  nucleosynthesis}.
\newblock {\em Astrophys. J.}, 488:515, 1997.

\bibitem{PhysRev.167.1170}
E.~R. Harrison.
\newblock Baryon inhomogeneity in the early universe.
\newblock {\em Phys. Rev.}, 167:1170--1175, Mar 1968.

\bibitem{Megevand:2004ry}
Ariel Megevand and Francisco Astorga.
\newblock {Generation of baryon inhomogeneities in the electroweak phase
  transition}.
\newblock {\em Phys. Rev. D}, 71:023502, 2005.

\bibitem{White:2021hwi}
Graham White, Lauren Pearce, Daniel Vagie, and Alexander Kusenko.
\newblock {Detectable Gravitational Wave Signals from Affleck-Dine
  Baryogenesis}.
\newblock {\em Phys. Rev. Lett.}, 127(18):181601, 2021.

\bibitem{Schaeffer1986}
R.~Schaeffer, P.~Delbourgo-Salvador, and J.~Audouze.
\newblock {INFLUENCE OF QUARK NUGGETS ON PRIMORDIAL NUCLEOSYNTHESIS}.
\newblock {\em Nature}, 317:407--409, 1985.

\bibitem{Alam:1998xb}
J.~Alam, B.~Sinha, and S.~Raha.
\newblock {Closing the universe with primordial quark nuggets}.
\newblock {\em Nucl. Phys. A}, 638:523--526, 1998.

\bibitem{Horvath:2008qj}
J.~E. Horvath.
\newblock {The search for Primordial Quark Nuggets among Near Earth Asteroids}.
\newblock {\em Astrophys. Space Sci.}, 315:361--364, 2008.

\bibitem{NANOGrav:2023hde}
Gabriella Agazie et~al.
\newblock {The NANOGrav 15 yr Data Set: Observations and Timing of 68
  Millisecond Pulsars}.
\newblock {\em Astrophys. J. Lett.}, 951(1):L9, 2023.

\bibitem{Lerambert-Potin:2021ohy}
Pauline Lerambert-Potin and Jos\'e~Antonio de~Freitas~Pacheco.
\newblock {Gravitational Waves from the Cosmological Quark-Hadron Phase
  Transition Revisited}.
\newblock {\em Universe}, 7(8):304, 2021.

\bibitem{Zic:2023gta}
Andrew Zic et~al.
\newblock {The Parkes Pulsar Timing Array third data release}.
\newblock {\em Publ. Astron. Soc. Austral.}, 40:e049, 2023.

\bibitem{Reardon:2023gzh}
Daniel~J. Reardon et~al.
\newblock {Search for an Isotropic Gravitational-wave Background with the
  Parkes Pulsar Timing Array}.
\newblock {\em Astrophys. J. Lett.}, 951(1):L6, 2023.

\bibitem{EPTA:2023sfo}
J.~Antoniadis et~al.
\newblock {The second data release from the European Pulsar Timing Array - I.
  The dataset and timing analysis}.
\newblock {\em Astron. Astrophys.}, 678:A48, 2023.

\bibitem{EPTA:2023fyk}
J.~Antoniadis et~al.
\newblock {The second data release from the European Pulsar Timing Array - III.
  Search for gravitational wave signals}.
\newblock {\em Astron. Astrophys.}, 678:A50, 2023.

\bibitem{Xu:2023wog}
Heng Xu et~al.
\newblock {Searching for the Nano-Hertz Stochastic Gravitational Wave
  Background with the Chinese Pulsar Timing Array Data Release I}.
\newblock {\em Res. Astron. Astrophys.}, 23(7):075024, 2023.

\bibitem{Sesana:2012ak}
A.~Sesana.
\newblock {Systematic investigation of the expected gravitational wave signal
  from supermassive black hole binaries in the pulsar timing band}.
\newblock {\em Mon. Not. Roy. Astron. Soc.}, 433:1, 2013.

\bibitem{XIAO20233158}
Yang Xiao, Jin~Min Yang, and Yang Zhang.
\newblock Implications of nano-hertz gravitational waves on electroweak phase
  transition in the singlet dark matter model.
\newblock {\em Science Bulletin}, 68(24):3158--3164, 2023.

\bibitem{Chen:2023bms}
Zu-Cheng Chen, Shou-Long Li, Puxun Wu, and Hongwei Yu.
\newblock {NANOGrav hints for first-order confinement-deconfinement phase
  transition in different QCD-matter scenarios}.
\newblock {\em Phys. Rev. D}, 109(4):043022, 2024.

\bibitem{He:2022amv}
Song He, Li~Li, Zhibin Li, and Shao-Jiang Wang.
\newblock {Gravitational waves and primordial black hole productions from
  gluodynamics by holography}.
\newblock {\em Sci. China Phys. Mech. Astron.}, 67(4):240411, 2024.

\bibitem{He:2023ado}
Song He, Li~Li, Sai Wang, and Shao-Jiang Wang.
\newblock {Constraints on holographic QCD phase transitions from PTA
  observations}.
\newblock 8 2023.

\bibitem{Chen:2022cgj}
Yidian Chen, Danning Li, and Mei Huang.
\newblock {Bubble nucleation and gravitational waves from holography in the
  probe approximation}.
\newblock {\em JHEP}, 07:225, 2023.

\bibitem{Morgante:2022zvc}
Enrico Morgante, Nicklas Ramberg, and Pedro Schwaller.
\newblock {Gravitational waves from dark SU(3) Yang-Mills theory}.
\newblock {\em Phys. Rev. D}, 107(3):036010, 2023.

\bibitem{Helmboldt:2019pan}
Alexander~J. Helmboldt, Jisuke Kubo, and Susan van~der Woude.
\newblock {Observational prospects for gravitational waves from hidden or dark
  chiral phase transitions}.
\newblock {\em Phys. Rev. D}, 100(5):055025, 2019.

\bibitem{Han:2023znh}
Xu~Han and Guoyun Shao.
\newblock {Stochastic gravitational waves produced by the first-order QCD phase
  transition}.
\newblock 12 2023.

\bibitem{Gao:2024pqm}
Fei Gao, Sichun Sun, and Graham White.
\newblock {A first-order deconfinement phase transition in the early universe
  and gravitational waves}.
\newblock 5 2024.

\bibitem{Caprini:2015zlo}
Chiara Caprini et~al.
\newblock {Science with the space-based interferometer eLISA. II: Gravitational
  waves from cosmological phase transitions}.
\newblock {\em JCAP}, 04:001, 2016.

\bibitem{RN8}
Sidney Coleman.
\newblock Fate of the false vacuum: Semiclassical theory.
\newblock {\em Physical Review D}, 15(10):2929--2936, 1977.

\bibitem{ellis2020gravitational}
John Ellis, Marek Lewicki, and Jos{\'e}~Miguel No.
\newblock Gravitational waves from first-order cosmological phase transitions:
  lifetime of the sound wave source.
\newblock {\em Journal of Cosmology and Astroparticle Physics}, 2020(07):050,
  2020.

\bibitem{RN20}
Pierre Binétruy, Alejandro Bohé, Chiara Caprini, and Jean-François Dufaux.
\newblock Cosmological backgrounds of gravitational waves and elisa/ngo: phase
  transitions, cosmic strings and other sources.
\newblock {\em Journal of Cosmology and Astroparticle Physics}, 2012:027, 2012.

\bibitem{RN15}
Astrid Eichhorn, Johannes Lumma, Jan~M. Pawlowski, Manuel Reichert, and
  Masatoshi Yamada.
\newblock Universal gravitational-wave signatures from heavy new physics in the
  electroweak sector.
\newblock {\em Journal of Cosmology and Astroparticle Physics}, 2021:006, 2021.

\bibitem{RN69}
Chiara Caprini, Mark Hindmarsh, Stephan Huber, Thomas Konstandin, Jonathan
  Kozaczuk, Germano Nardini, Jose~Miguel No, Antoine Petiteau, Pedro Schwaller,
  and Géraldine Servant.
\newblock Science with the space-based interferometer elisa. ii: Gravitational
  waves from cosmological phase transitions.
\newblock {\em Journal of cosmology and astroparticle physics}, 2016(04):001,
  2016.

\bibitem{espinosa2010energy}
Jose~R Espinosa, Thomas Konstandin, Jose~M No, and Geraldine Servant.
\newblock Energy budget of cosmological first-order phase transitions.
\newblock {\em Journal of Cosmology and Astroparticle Physics}, 2010(06):028,
  2010.

\bibitem{kamionkowski1994gravitational}
Marc Kamionkowski, Arthur Kosowsky, and Michael~S Turner.
\newblock Gravitational radiation from first-order phase transitions.
\newblock {\em Physical Review D}, 49(6):2837, 1994.

\bibitem{Zhou_2021}
Shuying Zhou, Song Shu, and Hong Mao.
\newblock Bubble dynamics in a strong first-order quark-hadron transition *.
\newblock {\em Chinese Physics C}, 45(4):043104, apr 2021.

\bibitem{Quiros:1999jp}
Mariano Quiros.
\newblock {Finite temperature field theory and phase transitions}.
\newblock In {\em {ICTP Summer School in High-Energy Physics and Cosmology}},
  pages 187--259, 1 1999.

\bibitem{Laine:2016hma}
Mikko Laine and Aleksi Vuorinen.
\newblock {\em {Basics of Thermal Field Theory}}, volume 925.
\newblock Springer, 2016.

\bibitem{Gell-Mann:1960mvl}
Murray Gell-Mann and M~Levy.
\newblock {The axial vector current in beta decay}.
\newblock {\em Nuovo Cim.}, 16:705, 1960.

\bibitem{Wang:2023omt}
Junrong Wang, Ziwan Yu, and Hong Mao.
\newblock {Bubble nucleation in the two-flavor quark-meson model*}.
\newblock {\em Chin. Phys. C}, 48(5):053105, 2024.

\bibitem{Shaojd2024long}
Jingdong Shao, Hong Mao, and Mei Huang.
\newblock Phase transition rate and gravitational wave spectra from first-order
  qcd phase transitions.
\newblock October 2024.

\bibitem{Bai:2018dxf}
Yang Bai, Andrew~J. Long, and Sida Lu.
\newblock {Dark Quark Nuggets}.
\newblock {\em Phys. Rev. D}, 99(5):055047, 2019.

\bibitem{PhysRevLett.132.221001}
Peter Athron, Andrew Fowlie, Chih-Ting Lu, Lachlan Morris, Lei Wu, Yongcheng
  Wu, and Zhongxiu Xu.
\newblock Can supercooled phase transitions explain the gravitational wave
  background observed by pulsar timing arrays?
\newblock {\em Phys. Rev. Lett.}, 132:221001, May 2024.

\bibitem{SALVIO2024138639}
Alberto Salvio.
\newblock Pulsar timing arrays and primordial black holes from a supercooled
  phase transition.
\newblock {\em Physics Letters B}, 852:138639, 2024.

\bibitem{2017EPJC77570K}
Archil {Kobakhidze}, Cyril {Lagger}, Adrian {Manning}, and Jason {Yue}.
\newblock {Gravitational waves from a supercooled electroweak phase transition
  and their detection with pulsar timing arrays}.
\newblock {\em European Physical Journal C}, 77(8):570, August 2017.

\bibitem{Jinno2019GravitationalWF}
Ryusuke Jinno, Hyeonseok Seong, Masahiro Takimoto, and Choong~Min Um.
\newblock Gravitational waves from first-order phase transitions:
  ultra-supercooled transitions and the fate of relativistic shocks.
\newblock {\em Journal of Cosmology and Astroparticle Physics}, 2019:033 --
  033, 2019.

\bibitem{Rajagopal:2000wf}
Krishna Rajagopal and Frank Wilczek.
\newblock {\em {The Condensed matter physics of QCD}}, pages 2061--2151.
\newblock 11 2000.

\bibitem{Huang:2004ik}
Mei Huang.
\newblock {Color superconductivity at moderate baryon density}.
\newblock {\em Int. J. Mod. Phys. E}, 14:675, 2005.

\bibitem{Alford:2007xm}
Mark~G. Alford, Andreas Schmitt, Krishna Rajagopal, and Thomas Sch\"afer.
\newblock {Color superconductivity in dense quark matter}.
\newblock {\em Rev. Mod. Phys.}, 80:1455--1515, 2008.

\bibitem{Li:2023yaj}
Yao-Yu Li, Chi Zhang, Ziwei Wang, Ming-Yang Cui, Yue-Lin~Sming Tsai, Qiang
  Yuan, and Yi-Zhong Fan.
\newblock {Primordial magnetic field as a common solution of nanohertz
  gravitational waves and the Hubble tension}.
\newblock {\em Phys. Rev. D}, 109(4):043538, 2024.

\end{thebibliography}
\end{document}